\documentclass[useAMS,usenatbib,usegraphicx]{mn2e}

\newcommand{\apj}{ApJ} 
\newcommand{\aap}{A\&A} 
\newcommand{\mnras}{MNRAS} 
\newcommand{\nat}{Nature}
\newcommand{\apjl}{ApJL}
\newcommand{\aj}{ApJ}
\newcommand{\apjs}{ApJS}
\newcommand{\pr}{PR}

\title[Estimating \textit{Herschel} redshifts]{H-ATLAS: Estimating redshifts of \textit{Herschel} sources from sub-mm fluxes}
\author[E. A. Pearson et al.]
{\parbox{\textwidth}
{\raggedright \Large E.A. Pearson,$^{1}$\thanks{E-mail: \texttt{Elizabeth.Pearson@astro.cf.ac.uk}}
S. Eales,$^{1}$
L. Dunne,$^{11}$      
J. Gonzalez-Nuevo,$^{13}$ 
S. Maddox,$^{12}$     
J.E. Aguirre,$^{2}$   
M. Baes,$^{3}$        
A. Baker,$^{4}$       
N. Bourne,$^{5}$      
C. M. Bradford,$^{6}$  
C. J. R. Clark,$^{1}$
A. Cooray,$^{7}$      
A. Dariush,$^{8}$     
G. De Zotti,$^{9,10}$ 
S. Dye,$^{5}$         
D. Frayer,$^{12}$     
H.L. Gomez,$^{1}$
A.I. Harris,$^{14}$   
R. Hopwood,$^{5}$     
E. Ibar,$^{15}$       
R.J. Ivison,$^{15}$   
M. Jarvis,$^{16}$     
M. Krips,$^{17}$      
A. Lapi,$^{18,9}$     
R.E. Lupu,$^{2}$     
M. J. Micha{\l}owski,$^{20}$
M. Rosenman,$^{2}$    
D. Scott,$^{21}$      
E. Valiante,$^{1}$
I. Valtchanov$^{22}$, 
P. van der Werf,$^{23}$ 
J.D. Vieira$^{24}$    
}\vspace{0.4cm}\\
\parbox{\textwidth}
{\raggedright \tiny $^{1}$School of Physics \& Astronomy, Cardiff University,The Parade, Cardiff, CF24 3AA, UK\\
$^{2}$Department of Physics and Astronomy, University of Pennsylvania, Philadelphia, PA 19104, USA\\
$^{3}$Sterrenkundig Observatorium, Universiteit Gent, Krijgslaan 281 S9,B-9000 Gent,Belgium\\
$^{4}$Department of Physics and Astronomy, Rutgers, The State University of New Jersey, Piscataway, NJ08854-8019, USA\\
$^{5}$School of Physics and Astronomy, University of Nottingham, Nottingham NG7 2RD, UK\\
$^{6}$Jet Propulsion Laboratory, M/S 169-506, 4800 Oak Grove Drive, Pasadena, CA 91109\\
$^{7}$Department of Physics and Astronomy, University of California, Irvine, 4186 Frederick Reines Hall, Irvine, CA 92617\\
$^{8}$Astrophysics Group, Imperial College London, Blackett Laboratory, Prince Consort Road, London SW7 2AZ, UK\\
$^{9}$SISSA, Via Bonomea 265, I-34136 Trieste, Italy\\
$^{10}$INAF-Osservatorio Astronomico di Padova, Vicolo dell’Osservatorio 5, I-35122 Padova, Italy\\
$^{11}$Department of Physics and Astronomy, University of Canterbury, Private Bag 4800, Christchurch, New Zealand\\
$^{12}$National Radio Astronomy Observatory, P.O. Box 2, Green Bank, WV 24944, USA\\
$^{13}$Instituto de Fisica de Cantabria, CSIC-UC, Av. de Los Castros s/n, Santander 39005, Spain\\
$^{14}$Department of Astronomy, University of Maryland, College Park, MD 20742, USA\\
$^{15}$UK Astronomy Technology Centre, Royal Observatory, Blackford Hill, Edinburgh EH9 3HJ, UK\\
$^{16}$Department of Physics, University of Oxford, Keble Road, Oxford, OX1 3RH\\
$^{17}$IRAM, Institut de Radio Astronomie Millimétrique, 300 rue de la Piscine, 38400 Saint-Martin-d'Hères, France\\
$^{18}$Dipartimento di Fisica, Universit\`{a} di Roma ‘Tor Vergata’, Via Ricerca Scientiﬁca 1, 00133 Roma, Italy\\
$^{20}$Scottish Universities Physics Alliance,Institute for Astronomy, University of Edinburgh, Royal Observatory, Edinburgh, EH9 3HJ\\
$^{21}$University of British Columbia, 6224 Agricultural Road, Vancouver, B.C. V6T 1Z1, Canada\\
$^{22}$European Space Agency SRE-SDH P.O. Box 78 ES 28691 Villanueva de la Cañada (Madrid) Spain\\
$^{23}$Leiden Observatory, Leiden University, PO Box 9513, 23100 RA Leiden, The
Netherlands\\
$^{24}$California Institute of Technology, 1216 E. California Blvd., Pasadena, CA 91125, USA\\
}}

\date{Released 2012}
\pagerange{\pageref{firstpage}--\pageref{lastpage}} \pubyear{2012}
\usepackage{amsmath}
\usepackage{amssymb}

\begin{document}

\label{firstpage}

\maketitle

\begin{abstract}
Upon its completion the \textit{Herschel} ATLAS (H-ATLAS) will be the largest submillimetre survey to date, detecting close to half-a-million sources. It will only be possible to measure spectroscopic redshifts for a small fraction of these sources. However, if the rest-frame spectral energy distribution (SED) of a typical H-ATLAS source is known, this SED and the observed \textit{Herschel} fluxes can be used to estimate the redshifts of the H-ATLAS sources without spectroscopic redshifts. In this paper, we use a subset of 40 H-ATLAS sources with previously measured redshifts in the range $0.5<z<4$.2 to derive a suitable average template for high redshift H-ATLAS sources. We find that a template with two dust components ($T_\textrm{c} = 23.9$ K, $T_\textrm{h} = 46.9$ K and ratio of mass of cold dust to mass of warm dust of 30.1) provides a good fit to the rest-frame fluxes of the sources in our calibration sample. We use a jackknife technique to estimate the accuracy of the redshifts estimated with this template, finding a root mean square of  $\Delta z/(1+z)$ = 0.26. For sources for which there is prior information that they lie at $z > 1$ we estimate that the rms of $\Delta z/(1+z)$ = 0.12. We have used this template to estimate the redshift distribution for the sources detected in the H-ATLAS equatorial fields, finding a bimodal distribution with a mean redshift of 1.2, 1.9 and 2.5 for 250, 350 and 500$\,\umu$m selected sources respectively.
\end{abstract}


\section{Introduction}
Much of the optical emission from distant galaxies is absorbed by dust and re-radiated at sub-millimeter (sub-mm) wavelengths \citep{Fixsen1998}.  Sub-mm observations have revealed a population of dusty galaxies at $z > 2$, previously hidden at optical wavelengths (see review by \citet{Blain2002}).  The inferred star formation rates for these galaxies are huge, averaging at $\simeq 400 \textrm{M}_{\bigodot}$ yr$^{-1}$ \citep{Coppin2008}.  Observations of sub-mm galaxies (SMGs) allow us to examine star formation in the early universe and the strong cosmic evolution in the star formation rate \citep{GLP2000}.  Ground based surveys have managed to identify and study individual sub-mm sources \citep{Barger1998, Hughes1998, Blain1999}.  Such surveys however covered small areas of sky and only found a few tens of SMGs and suffered from biases in their selections. The BLAST survey \citep{Devlin2009} covered $\sim 9\,$deg$^2$ of sky and found a few hundred SMGs \citep{EalesBLAST} but to really probe the evolution of the SMGs with redshift much larger blind surveys are needed.

In order to investigate the SMGs, particularly the evolution of the star formation rate and the luminosity function, we need to know the redshifts of all sources being considered. Ideally this is done by matching a source to an optical counterpart and then measuring the redshift of this counterpart spectroscopically. However the poor angular resolution of  sub-mm telescopes and high confusion between sources means that finding optical counterparts in this way is difficult.  One method to find counterparts is to first match the sub-mm source to a mid-IR or radio source, then match the mid-IR/radio source to its corresponding optical counterpart.  This can lead to a bias, however, as cold or high redshift objects are more likely to be undetected at mid-infrared and radio wavelengths \citep{Chapman2005, Younger2007}.

Fully exploiting the potential of sub-mm wavelengths on a large scale was impossible until the advent of the \textit{Herschel Space Observatory} \citep{Herschel}\footnote[1]{\textit{Herschel} is an ESA space observatory with science instruments provided by European-led Principal Investigator consortia and with important participation from NASA.}. The infrared emission of galaxies peaks between $70 - 500\,\umu$m, the wavebands that are covered by \textit{Herschel}'s two instruments: the Spectral and Photometric Imaging Receiver, SPIRE \citep{SPIRE}, and the Photodetector Array Camera and Spectrometer, PACS \citep{PACS}. The \textit{Herschel} Astrophysics Terahertz Large Area Survey, H-ATLAS \citep{HerAt}, covers 550 deg$^2$ of sky and is the largest sub-mm blind survey to date.  

The H-ATLAS fields were chosen partly due to the high quantity of complementary data at other wavelengths. However, less than 10\% of the H-ATLAS sources in the 15h field are detected by 	\textit{WISE} at 22$\,\umu$m \citep{Bond2012} and current large-area radio surveys only detect a tiny fraction of H-ATLAS sources. Nevertheless, \citet{Smith2011} and \citet{Fleuren2012} have shown that it is possible, using a sophisticated Baysian technique, to match the H-ATLAS sources to optically-detected galaxies directly. However, only approximately a third of the H-ATLAS sources have single reliable optical counterparts on images from the Sloan Digital Sky Survey (SDSS) \citep{Smith2011} which has limited subsequent investigations into the luminosity \citep{Dye2010} or dust mass \citep{Dunne2011} functions. Matching to the near infrared  images from the VIKING survey produces a higher proportion of counterparts, 51\% opposed to the 36\% provided by the optical \citep{Fleuren2012}, but there are still a large number of sources without counterparts.

CO line spectroscopy, using wide band instruments, can be used to accurately measure the redshift of sub-mm sources without the need for accurate optical positions \citep{Lupu2010, Frayer2011, Harris2011}. However, CO observations are time consuming and even with ALMA it will only be possible to measure redshifts for a tiny fraction of the H-ATLAS sources.

The only feasible method currently for estimating redshifts for such a large number of \textit{Herschel} sources is to estimate the redshifts from the sub-mm fluxes themselves. Previous attempts to estimate redshifts for \textit{Herschel} sources from the sub-mm fluxes have used as templates the spectral energy distributions (SEDs) of individual galaxies e.g. \citet{Lapi2011, GN2012}. Many of these template galaxies are at low redshift and their SEDs may not be representative of the SEDs of the high-redshift \textit{Herschel} sources and even if a high-redshift galaxy is used it may not be representative of the high-redshift population as a whole. For these reasons, we describe in this paper a method for creating a template directly from the sub-mm fluxes of all the high-$z$ H-ATLAS sources for which there are spectroscopic redshifts. The SEDs are also important for increasing our understanding of the population of high-redshift dusty galaxies and investigating the SEDs at the range of wavelengths in which the dust emission is at its peak. The average SED that we derive in this paper, although obviously telling us nothing about the diversity of the population, is still useful for comparing this population with dusty galaxies of low redshift \citep{Dunneeales2001, BBC2003}.

Section \ref{sec:Data} describes the observations on which the method is based.  We describe the method of template determination in Section \ref{sec:SED} and present the estimated redshift distributions in Section \ref{sec:zDist}.  We summarise our results in Section \ref{sec:Conc}. We assume $\Omega_\textrm{m} = 0.3$, $\Omega_{\lambda} = 0.7$, $H_0$ = 70 km s$^{-1}$ Mpc$^{-1}$.

\section{Data}
\label{sec:Data}
\subsection{FIR images and catalogues}

Phase 1 of the H-ATLAS survey covers around 160 deg$^2$ of sky with both PACS observations at 100 and 160$\,\umu$m and SPIRE observations at 250, 350 and 500$\,\umu$m.  However only a few percent of the H-ATLAS sources were detected at PACS wavelengths at greater than 5$\sigma$, so we have developed a method of estimating redshifts using only the SPIRE fluxes.  Phase 1 coincides with the three equatorial fields of the Galaxy and Mass Assembly, GAMA \citep{Driver2011}, spectroscopic survey. 

The FWHM beam sizes of the SPIRE observations are 18$''$, 25$''$ and 35$''$ for 250, 350 and 500$\,\umu$m respectively.  \citet{Pascale2011} describes the map-making procedure for the SPIRE observations. To find the sources, the MADX algorithm \citep{Maddox2010, Rigby2010} was used on the maps that had been passed through a point spread function filter.
The algorithm initially used the 250$\,\umu$m map to find the positions of sources detected above 2.5$\sigma$. The corresponding fluxes from the 350 and 500$\,\umu$m maps were then measured at these positions. If a source was detected at greater than 5$\sigma$ in any of the three wavebands then it was listed as a detection, with 78,014 sources extracted in total. The 5$\sigma$ sensitivities of the catalogues are 32, 36 and 45 mJy for 250, 350 and 500 $\umu$m, respectively.
The error on the flux, $\sigma_{\textrm{meas}}$, is the combined instrumental and confusion noise with an additional 7\% calibration error added in quadrature. The Phase 1 \textit{Herschel} maps and catalogues will be described fully in Valiante et al. (in prep.).

\subsection{Optical Counterparts}
\label{sec:OC}

The fields were chosen due to their lack of galactic cirrus (though G09 does still contain a large amount of cirrus) and large amount of complementary multi-wavelength data.  However the lack of radio and mid-IR data meant counterparts were found directly by applying a likelihood ratio technique \citep{Smith2011} to objects in the SDSS \citep{York2010} DR7 catalogue with a search radius of 10\arcsec. Only optical objects matched with a reliability factor R$\ge$80\% were considered as reliable matches.

\begin{figure}
	\centering
	\includegraphics[width = 0.35\textwidth, angle = 270]{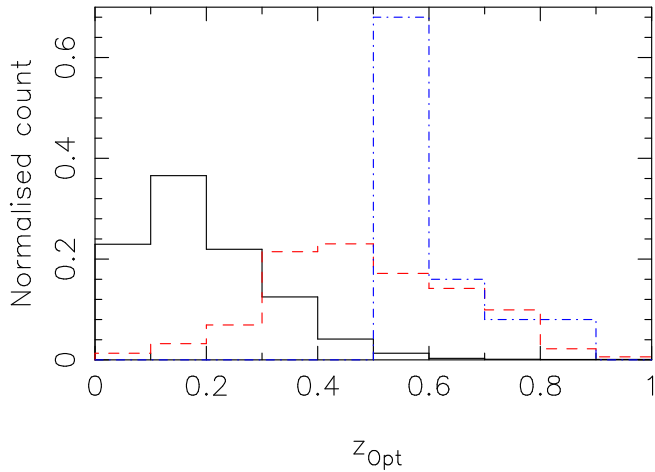}
	\caption[]{Redshift distributions of the H-ATLAS galaxies as determined from their SDSS counterparts. The solid black line shows those with measured spectroscopic redshifts and the dashed red line those with photometrically estimated redshifts only. The dot-dashed blue line shows the redshift distribution of the objects in the sample used to derive the template (Section \ref{sec:SED}): 25 spectroscopically observed sources with $0.5<z<1.0$ and $S_{250}>50$mJy.}

	\label{figure:zDistOpt}
\end{figure}

\begin{figure}
	\centering
	\includegraphics[width = 0.35\textwidth, angle = 270]{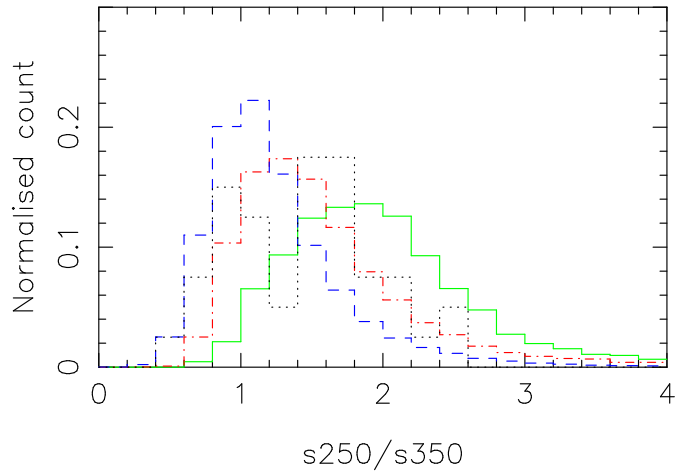}
	\caption[]{Histograms of the ratio of 250$\,\umu$m to 350$\,\umu$m fluxes. The solid green line represents those with spectroscopically measured optical counterparts. The dot-dashed red line shows sources with only photometric redshifts. The blue dashed line shows sources without any optical counterpart. The black dotted line shows the sample of 40 sources in the sample used to derive the template (Section \ref{sec:SED}). Sources without counterparts are redder in colour, indicating a higher redshift population.}
	\label{figure:SmithComp}
\end{figure}

23,312 sources have reliable optical counterparts. For these there is photometry in \textit{ugriz} and \textit{YJHK} from the SDSS and UKIDS Large Area Survey \citep{Lawrence2007}, respectively, and FUV and NUV data from GALEX \citep{Martin2005}.  12,136 sources also have spectroscopic redshifts available from the SDSS, 6dFGS \citep{Jones2009} and 2SLAQ-QSO/LRG \citep{Croom2009, Cannon2006} surveys and from the GAMA catalogues \citep{Driver2011}.  A further 10,972 photometric redshifts have been estimated from optical and near-IR photometry using the artificial neural network code (ANNz) \citep{Smith2011}.  These redshift distributions are shown in Fig \ref{figure:zDistOpt}. In Fig \ref{figure:SmithComp} sources without optical counterparts are shown to have slightly redder sub-mm colours, suggesting that they lie at higher redshifts than those with counterparts.

\subsection{CO Observations}
\label{sec:COObs}

We used fifteen H-ATLAS sources with redshifts from CO observations to construct our template. These sources are listed in Table \ref{table:HarrisZ}. Five of these are from \citet{Lupu2010}, who measured CO redshifts for sources
with $S_{500} > 100$mJy; seven are from \citet{Harris2011}, who observed galaxies whose sub-mm emission peaked at 350$\,\umu$m, indicating a high redshift; one is from \citet{Cox2011}, who studied one of the brightest sources in the GAMA 15hr field, which has the peak of its emission at 500$\,\umu$m; and the remaining two are as yet unpublished redshifts from the H-ATLAS team.

The selection criteria for these follow-up observations picked out bright galaxies that were likely to be at high redshift and so only represent the most luminous high-$z$ galaxies. The \textit{Herschel} colours of these galaxies are very red, which might introduce a bias towards colder objects. There is also a bias towards galaxies that are rich in CO gas, since not all sources observed in the CO programme were detected. Many of these sources are likely to have been strongly lensed \citep{Negrello2010, Harris2011}. As the gravitational magnification is likely to vary over a source it is possible that an unusually warm section of a galaxy might be magnified more strongly, boosting the flux at short wavelengths. However the dust detected at SPIRE wavelengths is likely to be cool and evenly distributed throughout the galaxy and so the \textit{Herschel} colours are likely to remain reasonably unaffected and resulting temperatures can be taken as safe upper limits.

\section{The Template}
\label{sec:SED}
\subsection{Sample selection}
To create the template we formed a sample of bright sources with accurately known redshifts. To do this we selected sources with either a redshift determined from the CO observations, $z_{\textrm{CO}}$, or an optically determined redshift, $z_{\textrm{spec}}$, with $0.5 \le z_{\textrm{spec}} < 1$.  In addition the flux must be greater than $50\,$mJy in at least one of the SPIRE wavelengths. Optically selected sources with $z_{\textrm{spec}} > 1$ are more likely to be quasars or atypical galaxies and so we did not use sources with optically determined redshifts above this reshift. The flux and redshift limits ensure we have a selection of high-$z$ sources for which we have accurate measurements of the SEDs.

We excluded sources at $z < 0.5$ for two reasons. First, these sources do not actually provide much extra information about the rest-frame \textit{Herschel} SEDs, because for low-redshift galaxies the SPIRE colours depend very weakly on dust temperature. Second, there is evidence from studies that combine the PACS and SPIRE data for individual sources \citep{Lapi2011,DSmith2012} and from stacking analyses (Eales et al. in prep.) that the SEDs of low-redshift and high-redshift Herschel sources are quite different.

These selection criteria produced a sample of 40 sources with known redshifts which are given in Table \ref{table:HarrisZ}: 15 sources with CO redshifts and 25 sources with optical redshifts.  There are actually many more sources in the redshift range $0.5 < z < 1.0$ with optical redshifts, but 25 were randomly chosen in order to prevent them from overwhelming the CO sources. We assume that this sample is representative of the whole survey; their redshifts and \textit{Herschel} colours are shown for comparison in Figs \ref{figure:zDistOpt} and \ref{figure:SmithComp}. The colours of this sample seem to be similar to those of sources with no optical counterpart. However, a possible bias may arise from the fact that all these sources are chosen to be bright and so will be among the most luminous H-ATLAS sources at their respective redshifts and so may not be representative of less luminous sources \citep{Casey2012etal}. We will use PACS data to test the dependence of dust temperature on luminosity in a later paper (Eales et al. in prep.).

\subsection{Creating the Template}
\label{sec:CtT}
We then transform these sources to their rest-frame wavelengths as determined by their $z_{\textrm{spec}}$ or $z_{\textrm{CO}}$, thus giving a range of flux measurements from $\sim50 - 350\,\umu$m. We then fit our model, based upon a modified black body spectrum, consisting of two dust components each with a different temperature:
\begin{eqnarray}
	S_{\nu} = A [ B_{\nu} (T_\textrm{h}) \nu ^{\beta} + a B_{\nu} (T_\textrm{c}) \nu ^{\beta} ]
\label{eqn:SEDfit}
\end{eqnarray}
where $S_{\nu}$ is the flux at a rest-frame frequency $\nu$, $A$ is a normalisation factor, $B_{\nu}$ is the Planck function, $\beta$ is the dust emissivity index, $T_\textrm{h}$ and $T_\textrm{c}$ are the temperatures of the hot and cold dust components, and $a$ is the ratio of the mass of cold dust to the mass of hot dust.

A two temperature model is important because galaxies with high far-infrared luminosities are known to contain a cold dust component \citep{Dunneeales2001}. We used $\beta = 2$ because recent \textit{Herschel} observations of nearby galaxies suggest this is a typical value \citep{Eales2012}. The SPIRE fluxes for the H-ATLAS sources do not give useful constraints on $\beta$ as they do not lie in the Rayleigh-Jeans region of the SED, where $\beta$ has the greatest effect.

For a given set of $T_\textrm{c}$, $T_\textrm{h}$ and $a$ the template was then fitted to the fluxes at their rest-frame wavelengths of all the sources within our sample. Different intrinsic brightnesses and distances caused a large variation in flux between sources and so we introduced an additional normalisation factor, $N_i$, for each source such that
\begin{eqnarray}
\label{eqn:chi}
\chi^2 = \sum\limits_{i=1}^{n} \left[\sum\limits^{\lambda} \frac{S_{\textrm{model},i} - N_i S_{\textrm{meas},i}}{N_i \sigma_{\textrm{meas},i}}\right]^2,
\end{eqnarray}
where $S_{\textrm{model},i}$ is the predicted flux of the $i^{\textrm{th}}$ source according to Equation \ref{eqn:SEDfit} for the set of values being considered and $S_{\textrm{meas},i}$ is the measured flux and $\sigma_{\textrm{meas},i}$ is the total error. For the $i^{\textrm{th}}$ source the measured fluxes and errors at all wavelengths are multiplied by $N_i$, and then the difference from the flux predicted by the model is found. Since the sources in our calibration sample are very bright, there are PACS measurements for many of them. In fitting the template, we used the PACS measurements for the sources as long as the rest-frame wavelength of the flux measurement was at $>$50$\,\umu$m; at shorter wavelengths
there is likely to be significant emission from dust that is not in thermal equilibrium. $\chi^2$ is a sum over all 40 sources in the sample and over all available wavelengths.

\begin{figure*}
	\centering
	\includegraphics[width = 0.65\textwidth, angle = 270]{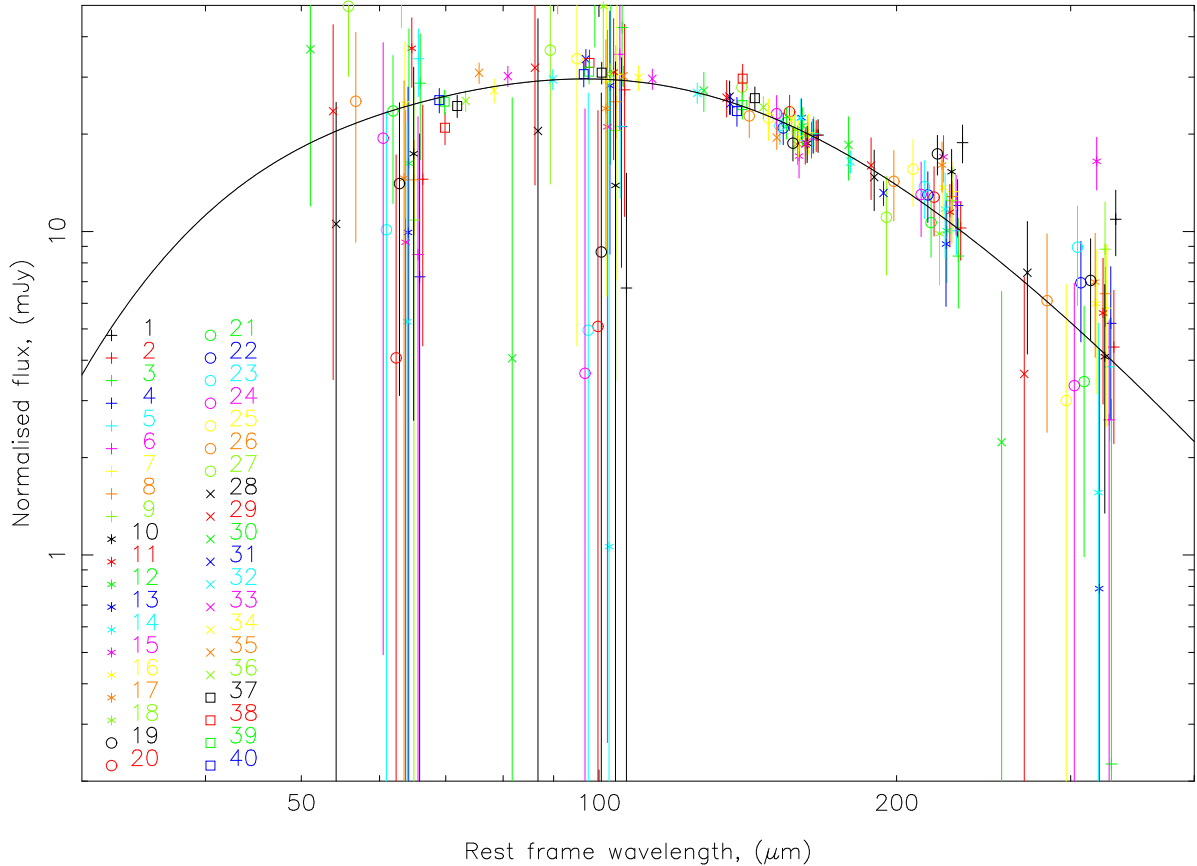}
	\caption[]{Best-fit model with the rest frame fluxes for all 40 of the sources in Table \ref{table:HarrisZ}, adjusted by their best normalisation factors, $N_i$. The red and blue lines show the SEDs for the individual dust components of our template.  All fluxes from a given source are shown with the same plot points, the key of which is given in Table \ref{table:HarrisZ}.}
	\label{figure:PlotTemplate}
\end{figure*}

\begin{figure}
	\centering
	\includegraphics[width = 0.35\textwidth, angle = 270]{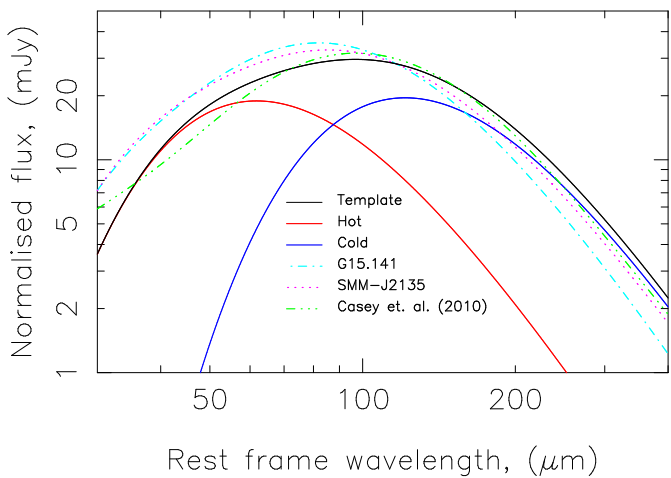}
	\caption[]{Best-fit model as compared with the SEDs from G15.141 (dotted magenta) and SMM J2135-0102 (dot-dashed cyan) used by \citet{Lapi2011} and the best fit SED from \citet{Casey2012etal} (green triple-dot dash). The comparative SEDs have been normalised to best fit the fluxes as they are shown in Figure \ref{figure:PlotTemplate}.}
	\label{figure:TemplateComp}
\end{figure}

For each combination of $T_\textrm{c}$, $T_\textrm{h}$ and $a$ we found the values of $N_i$ that gave the minimum value of $\chi^2$. Our best-fit model was the set of $T_\textrm{c}$, $T_\textrm{h}$ and $a$ that gave the lowest value of $\chi^2$ overall, resulting in the template shown in Figure \ref{figure:PlotTemplate} and the values given in Table \ref{table:JK}. Our best-fit model gives $T_\textrm{c} = 23.9$K, $T_\textrm{h} = 46.9$K with a ratio of cold to hot dust mass being 30.1. For comparison we have also shown the SEDs of SMM J2135-0102 ($z = 2.3$) and G15.141 ($z = 4.2$) in Figure \ref{figure:TemplateComp}, as used in \citet{Lapi2011} for estimating the redshifts of the sources in the H-ATLAS field observed during the \textit{Herschel} Science Demonstation Phase (SDP). All SEDs are normalised to the best values of $N_i$ given by our template as seen in Figure \ref{figure:PlotTemplate}. The template we find from the sample peaks at a slightly higher wavelength than that of those found in \citet{Lapi2011} though the Rayleigh-Jeans region has very similar slope, most likely as both use $\beta = 2$ for at least one of the dust components. When compared to the SED from \citet{Casey2012etal}, generated from spectroscopically selected HerMES galaxies, the peak lies in a very similar position. The SED dervied by \citet{Casey2012etal}  is controlled by a power law shortward of the peak to cover the mid-IR component, which is why is is so different from the other SEDs. However this region is well below the rest frame wavelength of sampled by our SPIRE observations.
 

It should be noted that the template is not expected to be a physically real SED but simply a statistical tool for estimating redshifts from SPIRE fluxes.  The peak of Fig \ref{figure:PlotTemplate} will represent the real SED of sources with $z \sim 2 - 4$, with the SED at longer wavelengths representing the real SED of H-ATLAS galaxies at lower redshift. In a later paper we will make a more detailed comparison of the SEDs of high-redshift H-ATLAS galaxies with low-redshift dusty galaxies. Here we note that the average SED is quite similar to the two-temperature SEDs found by \citet{Dunneeales2001} for luminous low-redshift dusty galaxies.

\subsection{A Jackknife Method for Testing the Template}
\label{sec:TT}
In order to test the accuracy of the redshifts determined from the template we used a jackknife technique.  From the initial selection of 40 sources we created two subsets by listing the sources by redshift and alternately placing them into each subset.  This ensured an even spread of redshifts and thus equal wavelength coverage. This was repeated twice more, this time splitting the sources randomly, resulting in three pairs of subsets from the initial data sample. For each subset we created a template as detailed in Section \ref{sec:CtT}. We then used the template to estimate the redshifts, $z_{\textrm{temp}}$, of the sources in the other sample from the pair. In estimating the redshifts the template was allowed to vary in redshift between $0 \le z < 20$ with the minimum  $\chi^2$ between the fluxes and the template giving the best estimate of $z_{\textrm{temp}}$. 

\begin{figure}
	\centering
	\includegraphics[width = 0.35\textwidth, angle = 270]{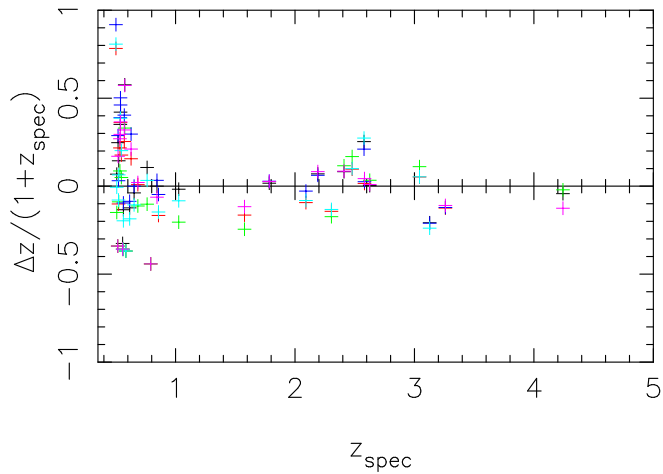}
	\caption[]{The data was split three ways into pairs of subsets. Each of these were used to create a template, then the template used to estimate the redshifts of the other subset in the pair. The resulting redhshift errors are shown here plotted against the spectroscopic redshifts. They key is given in Table \ref{table:JK}}
	\label{figure:ZVsZ}
\end{figure}

The temperatures and dust ratio values for the templates derived from the jackknife sets, as well as the values for the whole sample are shown in in Table \ref{table:JK}. To estimate the accuracy of the template derived from a set of sources, we calculate the value of 
\begin{eqnarray}
	\frac{\Delta z}{1 + z} \equiv \frac{z_{\textrm{temp}} - z_{\textrm{spec}}}{1 + z_{\textrm{spec}}}
\end{eqnarray}
for the sources in the other set from the pair (or the whole sample when the template is derived from the whole sample), where $z_{\textrm{spec}}$ is the best optical or CO redshift. Fig \ref{figure:ZVsZ} shows the estimates from all three jackknife pairs. The mean and root mean squared (rms) values for each template are shown in Table \ref{table:JK}. For comparison we have also used the two SEDs used in \citet{Lapi2011} to estimate the redshifts of the sources in our sample.

\begin{table*}
  \caption{Results of the jackknife tests applied to the data. `Template' indicates the subset used to create the template and the temperatures and dust mass ratios of the template are listed in the following three columns. `All' is the template resulting from using the whole sample and is the template that will be used in subsequent sections. The next two columns show our estimates of the redshift errors that will be obtained using that template, which were obtained by
comparing the redshift estimates and the spectroscopic redshifts for the sources in the other member of the jackknife pair (or all the sources for the template that was obtained from the whole sample). Column 5 shows the mean value of  
$\Delta z/(1+z_{spec})$ and column 6 gives the root mean squared (rms) of this. Column 7 gives the key for Fig \ref{figure:ZVsZ}. The two rows below the line show the result of testing two of the templates used by Lapi et al. (2011) against our calibration sample.}
  \centering
\label{table:JK}
  \begin{tabular}{ c c c c r c c }
  \hline
    Template &  $T_\textrm{c}$ & $T_\textrm{h}$ & $a$ & $\Delta z/(1+z)$ & rms & Key \\
  \hline
    1    & 24.8 & 45.5 & 22.25 &  0.06 $\pm$ 0.04 & 0.28 $\pm$ 0.03 & Black   \\
    2    & 22.2 & 43.0 & 22.22 & -0.03 $\pm$ 0.03 & 0.24 $\pm$ 0.03 & Red     \\
    3    & 18.8 & 39.6 & 20.97 & -0.06 $\pm$ 0.03 & 0.24 $\pm$ 0.03 & Green   \\
    4    & 26.6 & 51.1 & 44.55 &  0.08 $\pm$ 0.04 & 0.29 $\pm$ 0.03 & Blue    \\
    5    & 22.9 & 44.3 & 24.15 &  0.01 $\pm$ 0.03 & 0.25 $\pm$ 0.03 & Cyan    \\
    6    & 18.3 & 34.3 &  5.41 &  0.02 $\pm$ 0.04 & 0.27 $\pm$ 0.03 & Magenta \\
    All  & 23.9 & 46.9 & 30.10 &  0.03 $\pm$ 0.04 & 0.26 $\pm$ 0.03 &    -    \\
\hline
SMM & - & - & - & 0.135 & 0.332 & - \\
G15.141 & 32.0 & 60.0 & 50.0 & 0.269 & 0.431 & - \\
  \hline
  \end{tabular}
\end{table*}

As our estimate of the uncertainty in the redshifts estimates $z_{\textrm{temp}}$ from the template obtained from the whole sample, we use the average from all the jackknife tests in Table \ref{table:JK} giving a mean rms of $\Delta z / (1 + z) = 0.26$. Note that if we only look at sources where $z_{\textrm{spec}} > 1$ then the error is much less. Fig \ref{figure:ZVsZ} clearly shows that there is much higher accuracy above this cut off. If we restrict our error analysis to the sources in the template sample with $z_{\textrm{spec}} > 1$, we obtain a mean $\Delta z /(1 + z) = -0.013$ with and rms of 0.12. 

Our results are comparable to the error estimates given by \citet{Lapi2011}. When the templates from \citet{Lapi2011} (SMM J2135-0102 and G15.141) are used to estimate redshifts for our 40-source sample, there is a larger systematic error than when we use our own template, with the predicted redshifts considerably higher than the actual values. The reason for this can be seen in Fig \ref{figure:PlotTemplate}, which shows that the templates for SMM and G15.141 peak at lower wavelengths compared to our template.

For the subsequent sections we will use the template created when all sources in the sample were used (`All' in Table \ref{table:JK}). We have obtained this template from bright sources, whereas the majority of the Phase 1 sources have considerably lower signal to noise ratios, increasing the uncertainty in our redshift estimates. To gauge the total effect of this uncertainty on any particular redshift estimate we have used the template to estimate the redshifts for all the sources in the Phase 1 catalogue. We have then plotted the estimated redshifts against the statistical
error, which has been obtained by changing the redshift estimate until there is a change in $\chi^2 (\Delta \chi^2)$ of one (one `interesting' parameter, \citep{Avni1976}) (Fig \ref{figure:ZVsZerr}). This change in $\chi^2$ corresponds 
to a confidence region of 68\%. We can see that the uncertainty on $z$ grows with redshift up to $z = 2$, where it begins to fall again. 

The figure suggests that for a source that is detected at the signal-to-noise limit of the catalogue, the error is about 0.8 if the source is at a redshift of 3 but only 0.08 at a redshift of zero. This, however, ignores the important systematic error caused by the difference in dust temperature between low- and high-redshift H-ATLAS sources, which we address in the next section.

\subsection{Cold Sources at Low Redshift}

\begin{figure*}
	\centering
	\includegraphics[width = 0.65\textwidth, angle = 270]{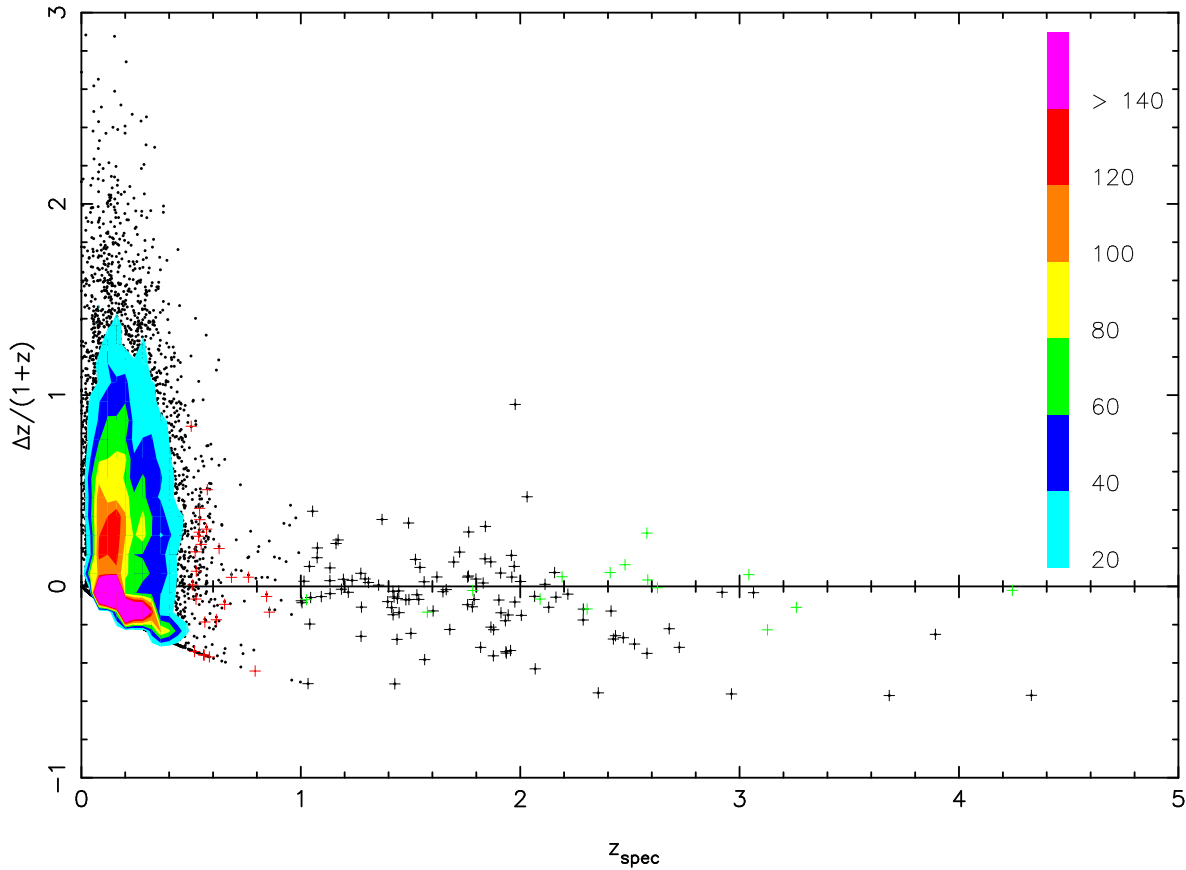}
	\caption[$z_{\textrm{spec}}$ against $\Delta z / (1+z)$ for all H-ATLAS sources.]{Plot of $z_{\textrm{spec}}$ against $\Delta z / (1+z)$ for all sources with measured redshifts, either CO redshifts or optical spectroscopy. Sources with $z_{\textrm{spec}} > 1$ are shown with crosses for clarity. Contours are included to show the density of sources at low redshifts. The key shows the number of sources in a bin where $\Delta z = 0.04$ and $\Delta(\Delta z / (1+z)) = 0.1$. Sources in red are the sources with optical redshifts that were used to create the template and the sources in green are the ones with CO measurements.}
	\label{figure:ZVsZ0to5}
\end{figure*}

\begin{figure}
	\centering
	\includegraphics[width = 0.35\textwidth, angle = 270]{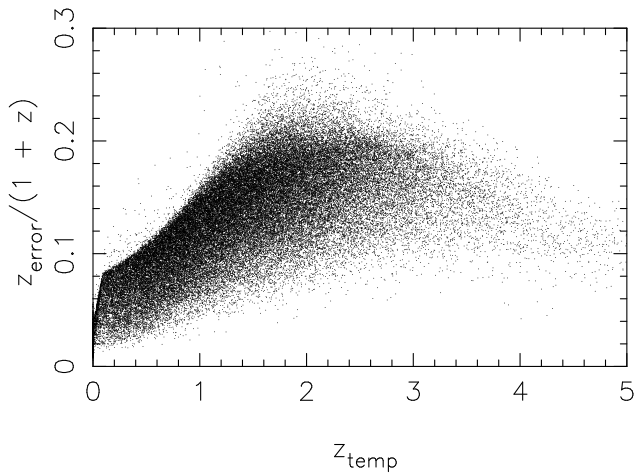}
	\caption[]{Plot of redshift according to our template against the estimated error as predicted from the $\chi^2$ corresponding to a confidence region of 68\% (see text). The hard edge at low $z_{\textrm{temp}}$ arises as these sources lie on the Rayleigh Jeans tail and are at the flux limit of the survey.}
	\label{figure:ZVsZerr}
\end{figure}

Fig \ref{figure:ZVsZerr} shows that the statistical error, $z_{err}$, for a redshift estimate for a low-redshift source is fairly small, but in reality there is a large systematic effect caused by the fact that low-redshift \textit{Herschel} sources have much cooler SEDs than the template we have derived from our high-redshift ($z > 0.5$) spectroscopic sample. This is shown dramatically in Fig \ref{figure:ZVsZ0to5}, where we have plotted $\Delta z / (1+z_{\textrm{spec}})$ for all H-ATLAS sources with either CO redshifts or optical counterparts (reliability $>$ 0.8) and spectroscopic redshifts. As expected, at $z >0.5$ the errors are quite small, but of the thousands of sources at $z < 0.5$ there are a large number with extremely large redshift discrepancies. As we demonstrate below this is likely to be mostly caused by a systematic temperature difference between low and high-$z$ \textit{Herschel} sources, but there will be some discrepancies due to gravitational lensing, in which the \textit{Herschel} source is really at a very high redshift with the apparent optical counterpart at much lower redshift being the graviational lens \citep{Negrello2010, GN2012}. The effect of this will be investigated in a subsequent paper.

We have investigated the possibility of systematic errors caused by temperature diffences by using a Monte-Carlo simulation. In this simulation we start with the Phase 1 H-ATLAS sources with reliable optical counterparts (reliability $>$ 0.8) and redshifts, either spectroscopic or estimates from optical photometry, $<0.4$. We then use these sources to generate probability distributions for the redshifts and the 250$\,\umu$m fluxes. The first step in the simulation is to create an artificial sample of galaxies by randomly drawing 250-$\,\umu$m fluxes and redshifts from these distributions. To produce an SED for each galaxy, we randomly assign one of the five average SEDs for low-redshift H-ATLAS galaxies from \citet{DSmith2012}. This library of SEDs seems the most appropriate for generating an artificial H-ATLAS sample, although we have also used 74 SEDs found for Virgo galaxies by \citet{Davies2012} and the 11 SEDs found for the KINGFISH sample by \citet{Galametz2012}, with very similar results. We use the SEDs and the redshifts to calculate 350$\,\umu$m and 500$\,\umu$m fluxes for each galaxy. The next step is to add noise to each galaxy. In order to allow for both instrumental noise and confusion, we add noise to each galaxy by randomly selecting positions on the real SPIRE images. We use the SPIRE images that have been convolved with the point spread function, since these were the ones used to find the sources and measure their fluxes. The final step in the simulation is to estimate the redshifts of the sources using our template.

\begin{figure}
        \centering
        \includegraphics[width = 0.35\textwidth]{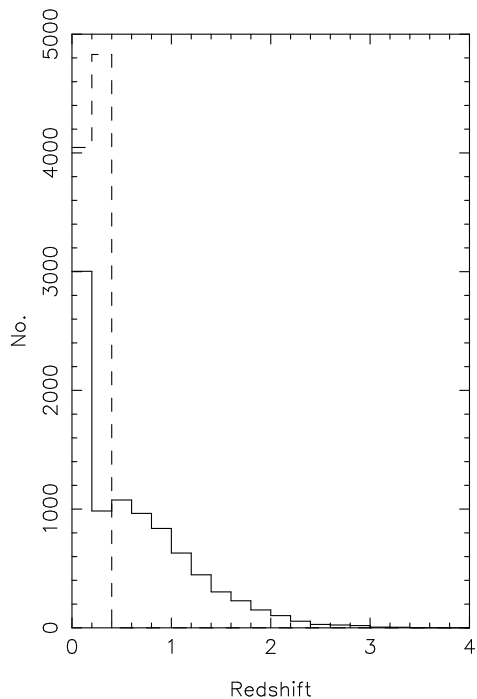}
        \caption[]{Results of Monte-Carlo simulation of our redshift estimation method for sources at low redshift, which are known to have cooler SEDs than our template. The dashed line shows the redshift distribution for sources in the Phase 1 catalogue with reliable identifications which have redshifts (spectroscopic or photometric) $<0.4$. The solid line shows the redshift distribution for these sources estimated using our template.}
        \label{figure:mc}
\end{figure}

Fig \ref{figure:mc} shows that the systematic errors can be very large. Although $\simeq$80\% of the sources have estimated redshifts $<1$, a significant fraction have higher estimated redshifts, although by $z>2$ the number of cool low-redshift sources that are spuriously placed at high redshift is very small. The simulation shows very clearly that one should not rely on this technique for estimating the redshifts of indvidual sources close to the flux limit of the survey. However, as we show in the next section, we can with care use it to draw some statistical conclusions about the survey.

\section{Redshift Distribution}
\label{sec:zDist}
\begin{figure*}
    \centering
	\includegraphics[width = 0.65\textwidth,trim = 0mm 45mm 0mm 0mm, clip=true]{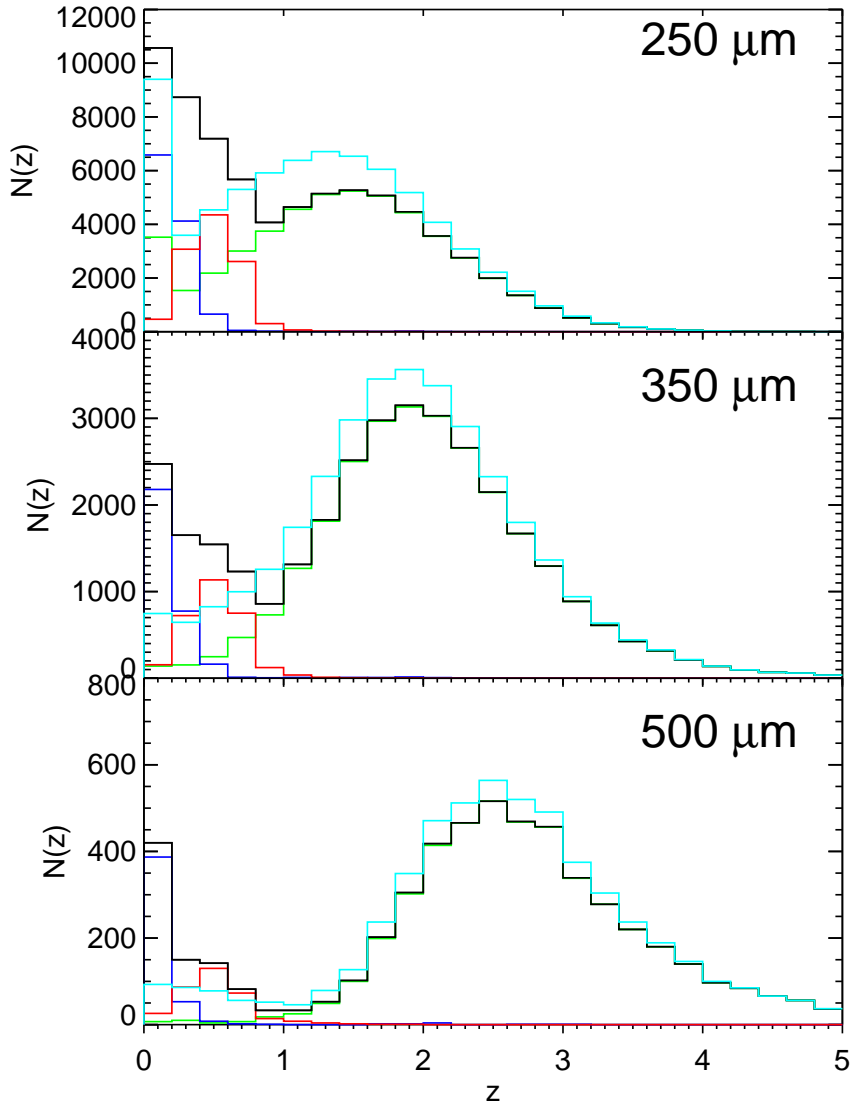}
	\caption[Redshift distributions of H-ATLAS sources]{Redshift distribution for sources with fluxes greater than 5$\sigma$ in the stated waveband. The upper plot shows the 250$\,\umu$m selection, with a median $z = $ 1.0, the middle 350$\,\umu$m with a median $z = $ 1.8 and the lower 500$\,\umu$m with a median $z = $ 2.5. All three show a large number of sources with $z < 0.2$ and a second broader distribution of sources at much higher redshifts. The dark blue line shows those sources with spectroscopic redshifts from optical counterparts. The red line shows those sources with optical photometric redshifts. The green line shows the redshifts estimated from the template for those sources with no reliable optical counterpart. The black line shows the sum of all three distributions (the median values stated are for these distributions). The light blue line shows the predicted redshift distributions if we do not use the redshifts of the optical counterparts but instead the redshifts estimated using the template.}
\label{figure:zDistIDL}
\end{figure*}

We used the following procedure to estimate the redshift distribution of the H-ATLAS sources. The template was used to estimate the redshifts, $z_{\textrm{temp}}$, of all the H-ATLAS Phase 1 sources without an optical counterpart, but where a reliable optical counterpart with a redshift was available we continued to use this value because of the problem described in the previous section. Fig \ref{figure:zDistIDL} shows the redshift distributions for sources with fluxes greater than 5$\sigma$ in a given band. The mean redshift increases with wavelength: $z = $1.2, 1.9 and 2.6 for 250, 350 and 500$\,\umu$m respectively due to the increasingly strong $K$-correction. A high-$z$ tail extends to $z \sim 5$ for 350 and 500$\,\umu$m selection and to $z \sim 4$ for 250$\,\umu$m.

We see a bimodal distribution with a large number of sources at low-$z$ ($z \le 0.8$), dominated by those sources with optical counterparts. This is seen in all three wavebands, though is most obvious at 250$\,\umu$m. By requiring that every source must have $z_{\textrm{temp}} \ge 0$, instrumental scatter may increase the size of the low-$z$ peak.  However most of the sources in the low-$z$ peak come from the optical counterparts and few of our estimated redshifts are used, particularly at longer wavelengths. Although there are undoubtedly H-ATLAS sources at low redshift that do not have reliable counterparts and which may be spuriously placed at high redshift, we do not see any way that this could create the bimodal redshift distribution seen for the 250-$\mu$m sample. We have also plotted in the figure the redshift distributions we obtain if we do not use the redshifts of the optical counterparts. At 250 $\mu$m, but not
at the other two wavelengths, there is still clear evidence of a bimodal distribution. The redshift distribution estimated by \citet{Dunlop2010} for the BLAST survey at 250$\,\umu$m is quite similar to ours and shows a similar bimodal distribution although it only contains a few tens of sources.

\begin{figure*}
    \centering
	\includegraphics[width = 0.65\textwidth,trim = 0mm 45mm 0mm 0mm, clip=true]{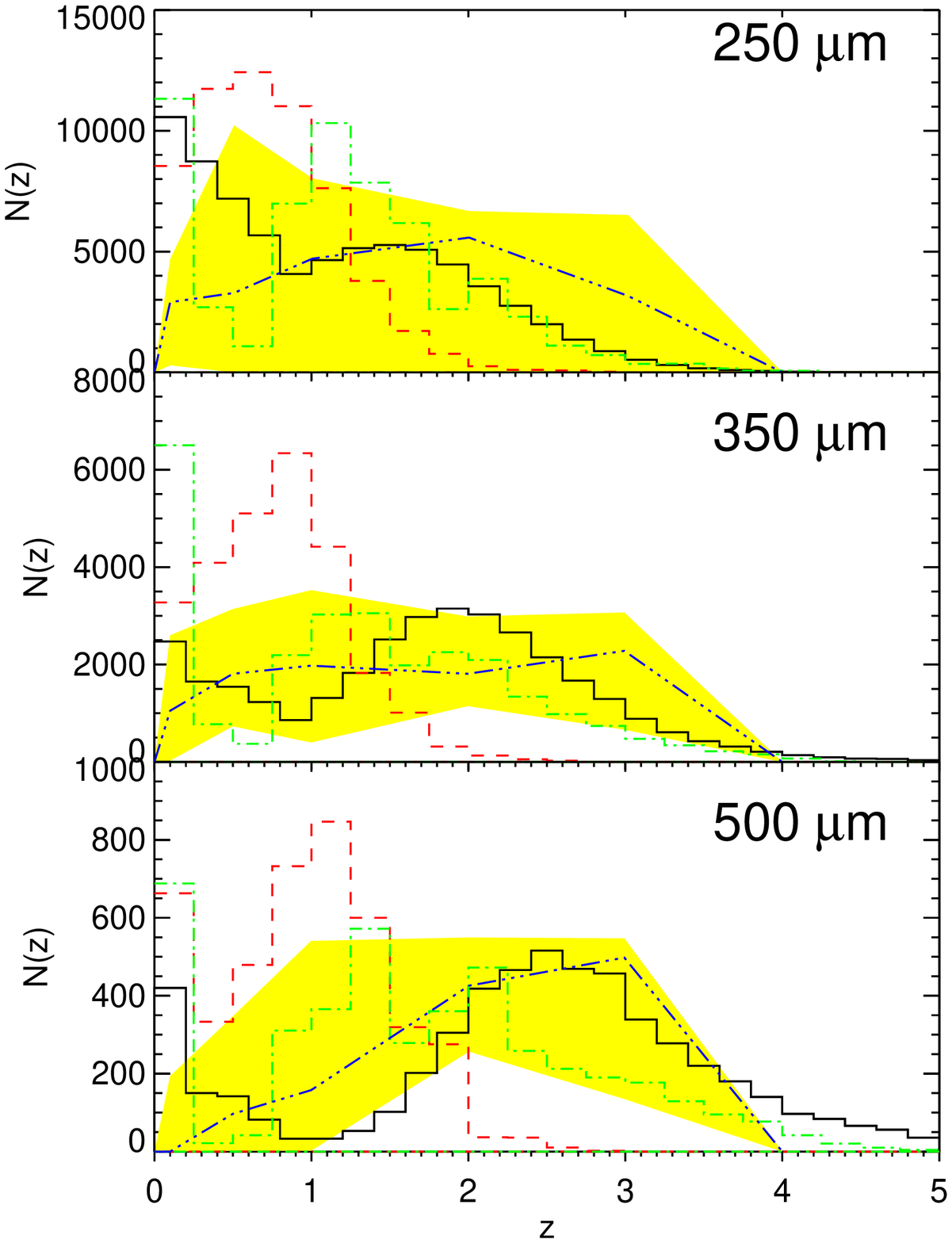}
	\caption[Redshift distribution comparison with \citet{HerAt}.]{Redshift distribution for sources with fluxes greater than 5$\sigma$ in the stated waveband. Overlaid are the models from \citet{HerAt}. The model from \citet{Lagache2004} is shown by the green dot-dashed line. The red dashed line is the SLUGS model. The blue dash-triple dotted line shows the model from \citet{MitchellWynne2012} with 1$\sigma$ confidence region in yellow. All models have been normalised to the number of sources detected with H-ATLAS.}
\label{figure:zDistIDLModels}
\end{figure*}

\citet{HerAt} presented  predicted H-ATLAS redshift distributions using models based on the SCUBA Local Universe and Galaxy Survey, SLUGS \citep{Dunne2000}, and the model described in \citet{Lagache2004}. The results are shown in Figure \ref{figure:zDistIDLModels} alongside our estimated distributions. The SLUGS model predicts few sources with $z > 2$, in strong disagreement with our results.  The \citet{Lagache2004} model predicts a bimodal distribution similar to what we find for the H-ATLAS sources and extends to redshifts similar to our distributions.  However our high-$z$ peaks are at a much higher redshift than predicted by the model.

\citet{Lagache2004} used both normal and starburst galaxies in their model. The differing cosmological evolution of these two populations causes the bimodal distribution seen in the model. Our redshift distribution also shows this bimodality suggesting that there really is two populations of galaxies, although  we cannot exclude the possibility that there is a single population, and the effects of the cosmic evolution of this population and the cosmological model combine to produce the bimodal redshift distrubution \citep{Blain1996}. This bimodality provides some support for the conclusions of \citet{Lapi2011} that the high-$z$ H-ATLAS sources represent a different population to the low-$z$ sources: spheroidal galaxies in the process of formation, rather than more normal star-forming galaxies seen at low redshift.

\citet{MitchellWynne2012} created a model by estimating the sub-mm redshift distribution from the strong cross-correlation of \textit{Herschel} sources with galaxy samples at other wavelengths, for which the redshift distribution is known. The initial redshift distributions were obtained by using 24$\,\umu$m Spitzer MIPS sources to cover the redshift range $0.5 < z < 3.5$ and optical SDSS galaxies to cover $0 < z < 0.7$. The authors estimate redshift distributions for samples of sources brighter than 20 mJy at the three SPIRE wavelengths, $\simeq$1.5-2 times fainter than the H-ATLAS limits. Their distributions agree quite well with the high-redshift peak of the H-ATLAS sources at all three wavelengths, but their distributions do not show the bimodal distribution that we find.

\begin{figure}
    \centering
	\includegraphics[width = 0.35\textwidth, angle = 270]{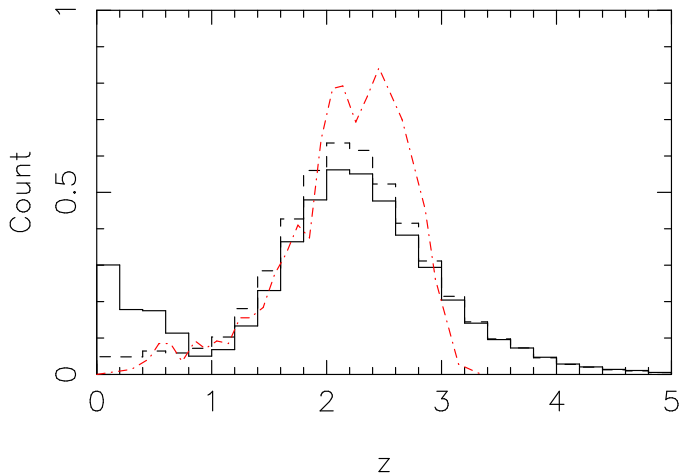}
	\caption[Comparison to Amblard and Lapi]{The estimated redshift distributions found by using our method and applying the cuts used by \citet{Amblard2010}: $S_{350} > 35$mJy, $S_{250}$ and $S_{500}> 3\sigma$. The solid black line shows our predicted redshift distribution if we use the redshifts of the reliable optical counterparts in preference to those estimated from the \textit{Herschel} fluxes. The black dashed line shows the results of using only the
redshifts estimated from the \textit{Herschel} fluxes. In the first case we find a mean redshift of $z=2.0$. The red dot-dashed line shows the redshift distribution obtained by Amblard et al. (2010).}
\label{figure:zDistA}
\end{figure}

\begin{figure}
    \centering
	\includegraphics[width = 0.35\textwidth, angle = 270]{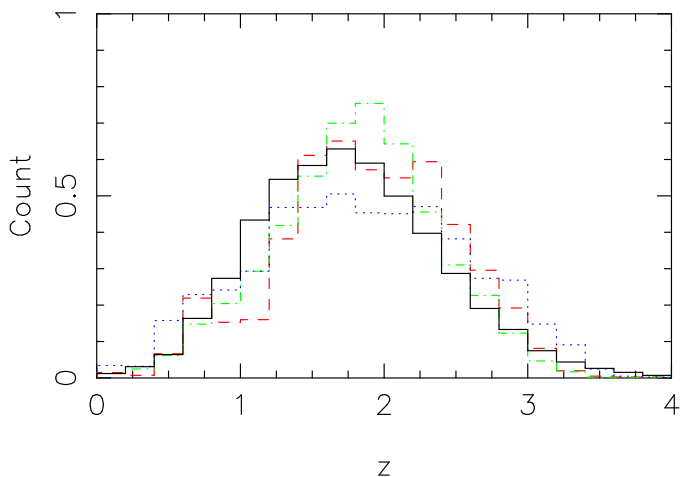}
	\caption[Comparison to Amblard and Lapi]{The estimated redshift distributions found by using our template and applying the cuts used by \citet{Lapi2011}: $S_{250} > 35$mJy, $S_{350} > 3\sigma$, no optical counterpart; solid black. The other lines shows the redshift distributions found by Lapi et al. (2011) for the H-ATLAS SDP field, the red dashed line with SMM J2135-0102 as the template, the green dot-dashed line with G15.141 as the template and the blue dotted line with Arp220 as
the template.}
\label{figure:zDistL}
\end{figure}

\citet{Amblard2010} and \citet{Lapi2011} have also estimated redshifts for H-ATLAS sources in the SDP field, which only contained $\sim 6000$ sources.  \citet{Amblard2010} used one-temperature modified black bodies with a range of temperature and $\beta$ to estimate the redshifts for sources from the SDP H-ATLAS field. These sources were selected to be detected at$> 3\sigma$ at 250 and 500$\,\umu$m and with fluxes greater than 35mJy (5$\sigma$) at 350$\,\umu$m.  These cuts bias against sources at lower redshifts, though the sample still includes several sources that were identified optically. 

\citet{Amblard2010} estimated a mean redshift of $z = 2.2$. In Fig \ref{figure:zDistA}, we have used our template to estimate redshifts for Phase 1 sources that satisfy the same flux criteria as used by \citet{Amblard2010}. Unlike \citet{Amblard2010}, we find a bimodal distribution, but it is worth noting that the majority of sources in the low-$z$ peak are redshifts from optical counterparts. We find many more sources beyond $z > 3$. This is presumably due to our use of a two-component dust model rather than the single-component model used by \citet{Amblard2010}. We find a mean redshift of 2.0, slightly lower than that found by \citet{Amblard2010}.

We also include in Figure \ref{figure:zDistA} our distribution of predicted redshifts if we now ignore the redshifts of any optical counterparts. In this case we see no low redshift peak and a mean $z = 2.3$ in good agreement with what \citet{Amblard2010} found. One possible explanation of the disappearance of the low-redshift peak are that these sources are mostly lensed high-redshift \textit{Herschel} sources. 

\citet{Lapi2011} used a S$_{250\,\umu m} > 35$mJy, S$_{350\,\umu m} > 3\sigma$ selection on SDP sources without an optical counterpart, again biasing against low-$z$ sources.  Three reference SEDs from galaxies at $z = 0.018, 2.3$ and $4.2$ were used to estimate redshifts from these fluxes and all produced similar distributions with a broad peak at $1.5 \lesssim z \lesssim 2.5$ and a tail up to $z \approx 3.5$.  Using our template and these same cuts, we find a mean of $z = 1.8$ (see Fig \ref{figure:zDistL}). Our and Lapi's estimates for the $z_{\textrm{temp}}$ distribution are very similar. This also confirms the methods of both \citet{Lapi2011} and \citet{GN2012} are reliable for estimating the redshifts of high-$z$ sources. \citet{Lapi2011} present a model for the formation of early-type galaxies that gives much better agreement with the estimated redshift distribution of H-ATLAS galaxies at $z > 1$.

\section{Conclusions}
\label{sec:Conc}
We generated a template for estimating the redshift of H-ATLAS galaxies using a sample of H-ATLAS galaxies with measured redshifts. Our best-fit template consists of two dust components with $T_\textrm{h} = 46.9$K, $T_\textrm{c} = 23.9$K, $\beta = 2$ and the ratio of cold dust mass to warm dust mass of 30.1. To estimate the uncertainty in the template we used a jackknife technique and found a mean $\Delta z / (1 + z) = 0.03$ with an rms of 0.26. If there is some \textit{a priori} knowledge that the source is at $z > 1$, we estimate a mean $\Delta z / (1 + z) = 0.013$ with an rms or 0.12.

This template was then used to estimate the redshifts of the entire H-ATLAS Phase 1 sources, though optical redshifts were used where available. Our redshift distributions show two peaks, suggesting there are two populations of sources experiencing different cosmological evolution.  The mean redshifts for sources detected at $> 5\sigma$ at three wavelengths are 1.2, 1.9 and 2.6 for 250, 350 and 500$\,\umu$m selected sources respectively.

\section*{Acknowledgments}
The \textit{Herschel}-ATLAS is a project with \textit{Herschel}, which is an ESA space observatory with science instruments provided by European-led Principal Investigator consortia and with important participation from NASA. The H-ATLAS website is http://www.h-atlas.org/.  GAMA is a joint European-Australasian project based around a spectroscopic campaign using the Anglo- Australian Telescope. The GAMA input catalogue is based on data taken from the Sloan Digital Sky Survey and the UKIRT Infrared Deep Sky Survey. Complementary imaging of the GAMA regions is being obtained by a number of independent survey programs including GALEX MIS, VST KIDS, VISTA VIKING, WISE, \textit{Herschel}-ATLAS, GMRT and ASKAP providing UV to radio coverage. GAMA is funded by the STFC (UK), the ARC (Australia), the AAO, and the participating institutions. The GAMA website is: http://www.gama-survey.org/. SPIRE has been developed by a consortium of institutes led by Cardiff University (UK) and including Univ. Lethbridge (Canada); NAOC (China); CEA, LAM (France); IFSI, Univ. Padua (Italy); IAC (Spain); Stockholm Observatory (Sweden); Imperial College London, RAL, UCL-MSSL, UKATC, Univ. Sussex (UK); and Caltech, JPL, NHSC, Univ. Colorado (USA). This development has been supported by national funding agencies: CSA (Canada); NAOC (China); CEA, CNES, CNRS (France); ASI (Italy); MCINN (Spain); SNSB (Sweden); STFC (UK); and NASA (USA). KSS is supported by the National Radio Astronomy Observatory, which is a facility of the National Science Foundation operated under cooperative agreement by Associated Universities, Inc. J.G.N. acknowledges ﬁnancial support from Spanish CSIC for a JAE-DOC fellowship and partial ﬁnancial support from the Spanish Ministerio de Ciencia e Innovacion project AYA2010-21766-C03-01.

\begin{table*}
  \caption{All sources used to make up the template sample.  Follow up observations were taken using the Caltech Submillimeter Observatory (CSO) with Z-Spec,  IRAM Plateau de Bure Interferometer (PdBI),  Green Bank Telescope (GBT) with Zpectrometer, Combined Array for Research in Millimeter-wave Astronomy (CARMA), Atacama Pathfinder Experiment (APEX), Sub Millimeter Array (SMA). All spectroscopic redshifts (those above the line) are from the Sloan Digital Sky Survey (SDSS) Data Release 7 (DR7) \citep{York2010}. CO redshifts are listed after the linebreak where H12 is\citet{Harris2011}, F11 is \citet{Frayer2011} and L12 is \citet{Lupu2010}.}
  \centering
  \begin{tabular}{ c l c c c}
  \hline
  \hline
    No. & H-ATLAS name & $z$ & Reference & Observations\\
  \hline
1   & HATLAS J143845.8+013504 & 0.501 & SDSS DR7 & SDSS  \\ 
2   & HATLAS J140746.5-010629 & 0.507 & SDSS DR7 & SDSS  \\ 
3   & HATLAS J090758.2-001448 & 0.516 & SDSS DR7 & SDSS  \\ 
4   & HATLAS J142534.0+023712 & 0.518 & SDSS DR7 & SDSS  \\ 
5   & HATLAS J143703.8+014128 & 0.522 & SDSS DR7 & SDSS  \\ 
6   & HATLAS J141815.6+010247 & 0.524 & SDSS DR7 & SDSS  \\ 
7   & HATLAS J083713.3+000035 & 0.534 & SDSS DR7 & SDSS  \\ 
8   & HATLAS J090359.6-004555 & 0.538 & SDSS DR7 & SDSS  \\ 
9   & HATLAS J140640.0-005951 & 0.539 & SDSS DR7 & SDSS  \\ 
10  & HATLAS J140930.6-013805 & 0.539 & SDSS DR7 & SDSS  \\ 
11  & HATLAS J141343.4+004041 & 0.546 & SDSS DR7 & SDSS  \\ 
12  & HATLAS J121353.8-024317 & 0.557 & SDSS DR7 & SDSS  \\ 
13  & HATLAS J092340.2+005736 & 0.560 & SDSS DR7 & SDSS  \\ 
14  & HATLAS J120248.3-022944 & 0.563 & SDSS DR7 & SDSS  \\ 
15  & HATLAS J114619.8-014356 & 0.571 & SDSS DR7 & SDSS  \\ 
16  & HATLAS J141429.0-000900 & 0.574 & SDSS DR7 & SDSS  \\ 
17  & HATLAS J085230.1+002844 & 0.584 & SDSS DR7 & SDSS  \\ 
18  & HATLAS J143858.1-010540 & 0.615 & SDSS DR7 & SDSS  \\ 
19  & HATLAS J084846.2+022032 & 0.627 & SDSS DR7 & SDSS  \\ 
20  & HATLAS J120246.0-005221 & 0.653 & SDSS DR7 & SDSS  \\ 
21  & HATLAS J113859.3-002934 & 0.684 & SDSS DR7 & SDSS  \\ 
22  & HATLAS J084217.0+010920 & 0.761 & SDSS DR7 & SDSS  \\ 
23  & HATLAS J090420.9+013038 & 0.792 & SDSS DR7 & SDSS  \\ 
24  & HATLAS J114023.0-001043 & 0.844 & SDSS DR7 & SDSS  \\ 
25  & HATLAS J141148.9-011439 & 0.857 & SDSS DR7 & SDSS  \\
  \hline

26  & HATLAS J142935.3-002836 & 1.026 & & ZSpec, CARMA\\
27  & HATLAS J090740.0-004200 & 1.577 & L12  & CSO \\
28  & HATLAS J091043.1-000321 & 1.784 & L12  & CSO \\
29  & HATLAS J085358.9+015537 & 2.091 & & ZSpec, PdBI\\
30  & HATLAS J115820.2-013753 & 2.191 & H12 & GBT\\ 
31  & HATLAS J090302.9-014127 & 2.308 & L12  & CSO, CARMA, GBT, PdBI \\
32  & HATLAS J084933.4+021443 & 2.410 & H12 & CARMA, GBT\\
33  & HATLAS J141351.9-000026 & 2.478 & H12 & GBT\\ 
34  & HATLAS J113243.1-005108 & 2.578 & H12 & GBT\\ 
35  & HATLAS J091840.8+023047 & 2.581 & H12 & GBT\\
36  & HATLAS J091305.0-005343 & 2.626 & L12, F11       & CSO, GBT, PdBI\\
37  & HATLAS J090311.6+003906 & 3.037 & L12, F11, H12  & CSO, PdBI, GBT\\
38  & HATLAS J113526.3-014605 & 3.128 & H12 & GBT\\ 
39  & HATLAS J114637.9-001132 & 3.259 & H12 & GBT\\ 
40  & HATLAS J142413.9+022303 & 4.243 & \citet{Cox2011}    & APEX, FLASH+, PdBI, SMA\\

  \hline
  \label{table:HarrisZ}
  \end{tabular}
\end{table*}

\bibliographystyle{apj}

\label{lastpage}

\end{document}